# Monte Carlo investigation into feasibility and dosimetry of flat Flattening Filter Free beams

**Sergei Zavgorodni[1,2]**

[1] *Radiation Therapy Program, Dept of Medical Physics, BC Cancer Agency – Vancouver Island Centre*

5   [2]*University of Victoria, Department of Physics & Astronomy*

Email: szavgorodni@bccancer.bc.ca

## Abstract

10   Flattening filter free (FFF) beams due to their non-uniformity, are sub-optimal for larger field sizes. The purpose of this study was to investigate the incident electron beam distributions that would produce flat FFF beams without the use of flattening filter. Monte Carlo (MC) simulations with BEAMnrc and DOSXYZnrc codes have been performed to evaluate the feasibility of this approach. The dose distributions in water for open 6MV beams were simulated using Varian 21EX linac head model, which will be called flattening filter (FF) model. Flattening filter has then been removed from FF model, and MC simulations were performed using (1) 6 MeV electrons incident on the target, (2) 6 MeV electron beam with electron angular distributions optimized to provide as flat dose profiles as possible. Configuration (1) represents FFF beam while configuration (2) allowed producing a flat FFF (F4) beam. Optimizations have also been performed to produce flattest profiles for a set of dose rates (DRs) in the range from 1.25 to 2.4 of the DR of FF beam. Profiles and percentage depth doses PDDs from 6MV F4 beams have been calculated and compared to those from FF beam. Calculated profiles demonstrated improved flatness of the FFF beams. In fact, up to field sizes within the circle of 35 cm diameter the flatness of F4 beam at $d_{max}$ was better or comparable to that of FF beam. At 20 cm off-axis the dose increased from 52% for FFF to 92% for F4 beam. Also, profiles of F4 beams did not change considerably with depth, and for large fields out-of-field dose was reduced by about a factor of two compared to FF beam. PDDs from F4 beams were similar to those of FFF beam. The DR for the largest modeled (44 cm diameter) F4 beam was higher than the DR from FF beam by a factor of 1.25. It was shown that the DR can be increased while maintaining beam flatness, but at the cost of reduced field size.

## Introduction

30   Recent research and introduction of the flattening filter free (FFF) beams to clinical practice generated considerable interest in radiotherapy with most publications highlighting advantages of FFF beams as compared to traditional FF beams. These include increased efficiency of the dose delivery, reduced out-of-field dose by about a factor of two, over 70% reduced leakage dose in the patient plane (Titt, Vassiliev *et al.*, 2006a, Titt, Vassiliev *et al.*, 2006b, Vassiliev, Titt *et al.*, 2006, Vassiliev, Kry *et al.*, 2007, Kry and

35   et al., 2010, Duan, Spencer *et al.*, 2011, Georg, Knoos *et al.*, 2011, Stevens, Rosser *et al.*, 2011, Almberg, Frengen *et al.*, 2012, Dzierma, Licht *et al.*, 2012).

These benefits result from removing attenuation of the primary beam as well as removing the scatter radiation component originating from the flattening filter. Reduction of the head scatter improves dosimetry of the FFF beams resulting in reduction of output variation with radiation field size as well as 40 reduced variation of all field size dependent parameters (Ponisch, Titt *et al.*, 2006, Zhu, Kang *et al.*, 2006, Kragl, af Wetterstedt *et al.*, 2009, Dalaryd, Kragl *et al.*, 2010). Lower value of leaf transmission as well as less transmission variation off axis was reported (Hrbacek, Lang *et al.*, 2011). It has also been reported that FFF beams will be easier to model in treatment planning algorithms potentially resulting in increased accuracy of the dose calculations (Cashmore, 2008, Georg, Knoos *et al.*, 2011, Kragl, Albrich *et al.*, 2011,
45 Cashmore, Golubev *et al.*, 2012) or improved calculation speed.

In addition, FFF beams demonstrated far less dosimetric sensitivity to variations in the beam steering (Cashmore, 2008). In experiments reported by Cashmore et al artificially introduced steering currents did not change the shape of the beam but only slightly shifted the beam position, while in traditional FF beams the same changes resulted in considerable asymmetry of the delivered dose.

50 Treatment planning studies have been reported using FFF beams to treat various treatment sites such as prostate (Vassiliev, Kry *et al.*, 2007, Zwahlen, Lang *et al.*, 2012) , lung (Stevens, Rosser *et al.*, 2011, Navarria, Pentimalli *et al.*, 2012, Verbakel, Ong *et al.*, 2012), breast (Stevens, Rosser *et al.*, 2011, Thomas, Popple *et al.*, 2012, Xie, Cao *et al.*, 2012), larynx (Stevens, Rosser *et al.*, 2011), head-and-neck (Abacioglu, Unal *et al.*, 2012, Wu, Yoo *et al.*, 2012), pancreas (Atwood, Mok *et al.*, 2011), liver
55 (Arcangeli, Castiglioni *et al.*, 2012), brain (Kim, Rice *et al.*, 2012, Thomas, Popple *et al.*, 2012), spine (Thomas, Popple *et al.*, 2012, Verbakel, Ong *et al.*, 2012), esophagus (Nicolini, Ghosh-Laskar *et al.*, 2012), cervix (Thomas, Popple *et al.*, 2012). Most often FFF beams were used with stereotactic body radiotherapy (SBRT) technique, though use of these beams with IMRT and VMAT have also been evaluated. These studies report that for smaller volumes FFF beams demonstrated their benefits providing
60 adequate PTV coverage combined with increased treatment efficiency.

However, considerable limitations associated with clinical use of FFF beams have also been reported. With larger PTVs, it proved increasingly difficult to obtain adequate coverage with FFF beams, and the plans were generally under dosing peripheral regions of larger lung PTVs and breast plans (Stevens, Rosser *et al.*, 2011). Use of FFF also resulted in increased MUs for IMRT and VMAT plans (Duan,
65 Spencer *et al.*, 2011, Zhang, Ku *et al.*, 2011, Almberg, Frengen *et al.*, 2012, Cashmore, Golubev *et al.*, 2012). This is because medium and large un-flattened fields require extra modulation in order to achieve required dose uniformity. Simple conformal plans, if re-planned for FFF beams, would require use of IMRT techniques resulting in higher MUs and also necessitating extra work load associated with these techniques in clinical setting.

70 These limitations of FFF beams inspired investigations into improving their flatness. Chofor et al (Chofor, Harder *et al.*, 2011) proposed "direction-selective filter" (DSF) to partly compensate for shortcomings of existing FFF. As a matter of fact "direction-selective filter" is a re-designed flattening filter that has been moved up to the level of target assembly. The DSF allowed achieving flatness for the fields with diagonal size not exceeding 15 cm in any direction, which includes fields such as 10x10 cm$^2$,
75 4x14.5 cm$^2$ or less.

Tsiamas et al studied effect of electron angular spread (Tsiamas, Seco *et al.*, 2011) on the dose profile flatness. They performed Monte Carlo simulations where angular spread of the electrons incident on the target was set to 5 and 10 degrees. This resulted in improved beam flatness and flattening of photon beam for the field sizes between 10 x10 cm$^2$ and 15 x15 cm$^2$ and partial flattening for field sizes from 15 x15
80 cm$^2$ to 30 x30 cm$^2$.

This paper aims to determine the optimal angular distribution of incident electrons that would produce as flat dose distribution as possible without the use of flattening filter. It also investigates improvement of beam flatness as well as dose rates that could be achieved from such distribution of incident electrons and evaluates of basic dosimetry of achieved beams.

85 ## Materials and Methods

### Terminology

The following terminology will be used through the paper: flattening filter free (FFF) beam, beam with flattening filter present will be referred to as flattening filer (FF) beam, relative dose rate (RDR), dose rate relative to FF beam will be referred to as $RDR_{FF}$, dose rate relative to FFF beam will be referred to as $RDR_{FFF}$, flat FFF will be used interchangeably with F4 beam, "convergence angle" of the incident electron beam is defined in Figure 2 as angle $\theta$; angular distribution of incident electrons will imply distribution of convergence angles as cylindrical symmetry of the beam is assumed.

### Monte Carlo modeling

For Monte Carlo (MC) modeling of the radiation transport through the linac head into water BEAMnrc/DOSXYZnrc package (Rogers, Walters *et al.*, 2009) was used within Vancouver Island Monte Carlo (VIMC) system (Zavgorodni, Bush *et al.*, 2007, Bush, Townson *et al.*, 2008) environment. DOSXYZnrc code used BEAMnrc as a shared library particle source utilizing directional bremsstrahlung splitting (Kawrakow, Rogers *et al.*, 2004) for improved efficiency. Our model of a standard 21EX linac head (schematically shown in Figure 1a) has been well benchmarked previously to produce excellent agreement with measured data (Zavgorodni, Locke *et al.*, 2005, Gagné and Zavgorodni, 2007, Bush, Zavgorodni *et al.*, 2009, Basran, Zavgorodni *et al.*, 2010, Bush, Gagné *et al.*, 2011). For this project modifications to the shielding geometry (not shown in Figure 1) have been added to the model to improve agreement of measured and calculated dose profiles for 40x40cm² fields along diagonal directions. This model will be called Flattening Filter (FF) model further in this paper.

In order to produce MC model for FFF beam, the flattening filter has been removed from the FF linac model while all other model components remained intact. MC simulations were then performed with 1) FFF model using electrons with the same 6 MeV beam parameters as in FF model, 2) convergent 6 MeV electron beams (Figure 1). Figure 1b represents flattening filter free (FFF) beam while Figure 1c shows FFF model with convergent beam of incident electrons. It is expected that, unlike FF beams, radial spread of the incident electron beams will not be an important factor in forming the shape of dose profile (Cashmore, 2008). In this paper all electrons were converging into a circle of 1mm radius on the target.

The dose distributions were calculated in 60x60x40 cm³ water phantom with voxel sizes variable in all directions from 0.2 cm to 1 cm with smaller dimensions being along the regions with expected steep dose gradients such as build up region and beam penumbra. The source to surface distance (SSD) of 90cm was used in the calculations and dose uncertainty of less than 0.5% achieved for in-field voxels at $d_{max}$.

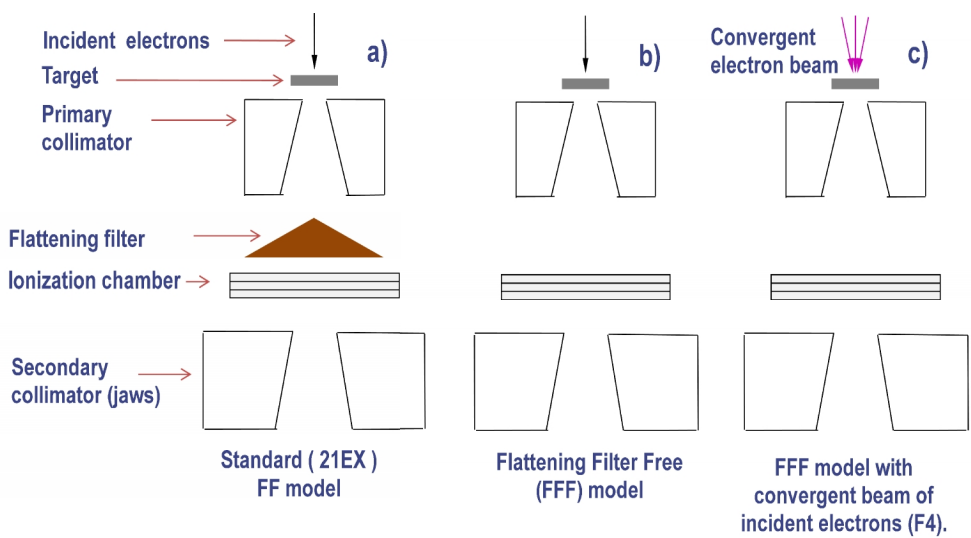

Figure 1. Diagram demonstrating geometries of a) standard 6MV beam with flattening filter (FF beam), b) geometry of FFF model used in this paper, and c) geometry of FFF beam combined with convergent electron beam incident on the target.

### Angular distribution of electrons, incident on the target, to produce flat FFF beam

In order to determine optimal angular distribution of electrons to produce flat diagonal dose profile for 40x40 cm$^2$ field in the geometry of Figure 1c, we modeled a set of dose distributions generated by convergent electron beams with different angles of convergence $\theta_i$ (Figure 2). For every angle of convergence $\theta_i$ the electrons entered the target within a narrow ($1^0$ degree) solid angle. Twenty one angles of convergence $\theta_i$ in the range from $0^0$ to $20^0$ were used providing 21 dose distributions. Diagonal profiles $D_i(r)$, normalized at the central axis, as well as the dose per incident electron at the central axis $D_i(r_0)$ (Gy/e$^-$) were extracted from these dose distributions and stored to be used by optimization algorithm.

Simulated annealing optimization algorithm was then employed to optimize the weights $w_i$ of the scored dose distributions to receive as flat profile $D_{opt}(r)$ as possible while maintaining highest possible dose per incident electron. The optimal angular distribution of the incident convergent electrons then corresponds to the distribution of optimized weights $w_i$. Given that we are aiming for a perfectly flat dose profile, which implies relative dose off-axis being a unit, objective function to be minimized becomes

$$F_{obj} = \frac{\sum_{i=1}^{N} w_i \sum_{j=1}^{M} \left(1 - D(r_j)\right)^2}{\sum_{i=1}^{N} w_i},$$

Where $N$ is the number of convergence angles and $M$ is the number of radial of the dose profile points used in optimization. Boundary condition is applied for the RDR$_{FF}$ to be set greater than required minimum $RDR_{FF} > RDR_{min}$.

These optimizations were repeated for a set of $RDR_{min}$ values in the range between RDR$_{FF}$ and RDR$_{FFF}$. For each $RDR_{min}$ value the radius of 90% dose level within $D_{opt}(r)$ was found and used to quantify the size of flat area of the dose distribution. This value of the dose level was chosen as an indication of the field size that would provide +/- 5% dose uniformity in the high dose region.

Similar approach has been previously used by our group in order to optimize the shape of electron profiles (Bush, Zavgorodni *et al.*, 2009) that provided best match of modeled and measured dose distributions in water.

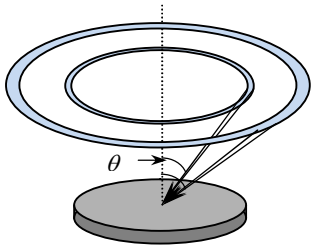

Figure 2. Electron beams incident on a target within small solid angles with different convergence angles $\theta_t$. Two out of modeled twenty one solid angles are shown.

### Comparing characteristics of F4 beams to those of FF and FFF beams.

Using the optimal angular distribution of electrons on the target, that provided flattest diagonal profile for 40 x 40 cm$^2$ field, the dose distributions in water were calculated for 4x4, 10x10, 20x20 and 40x40 cm$^2$ fields, and transverse profiles as well as PDDs were extracted for comparison with profiles from standard FF beam.

## Results

### Validation of MC model

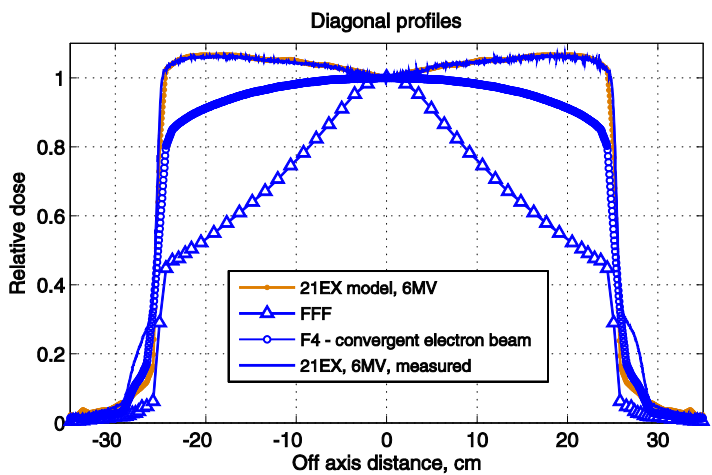

155

Figure 3. Diagonal dose profiles in water calculated at d$_{max}$ using our MC model of 21EX linac for open 40x40 cm$^2$ 6MV beam, open 40x40 cm$^2$ 6MV FFF beam, open 40x40 cm$^2$ 6MV F4 beam. Dose profile measured using IC13 ionization chamber for 21EX Varian linac is shown for comparison. All profiles are normalized at the beam central axis.

160 Figure 3 shows measured and MC calculated diagonal dose profiles for 6MV 40x40 cm$^2$ field. Unsmoothed measured data were used deliberately to demonstrate that agreement of calculated profiles with measurement is well within combined measurement and statistical uncertainties. Note that large field profiles are notoriously difficult to match to experimental data (Chibani, Moftah *et al.*, 2011). Diagonal profiles of 40x40 cm$^2$ are the largest possible profiles capturing considerably larger fraction of particle
165 fluence than any other measurable profile. These profiles are not commonly shown in MC model validation results potentially due to increasing difficulty of obtaining satisfactory agreement with measured data for such large fields. This figure shows excellent agreement of MC and measured profiles for our 21 EX model that was used as a basis for FFF and F4 modeling in this paper. Also shown in this figure are profiles for FFF and flattest achieved (F4) beams.

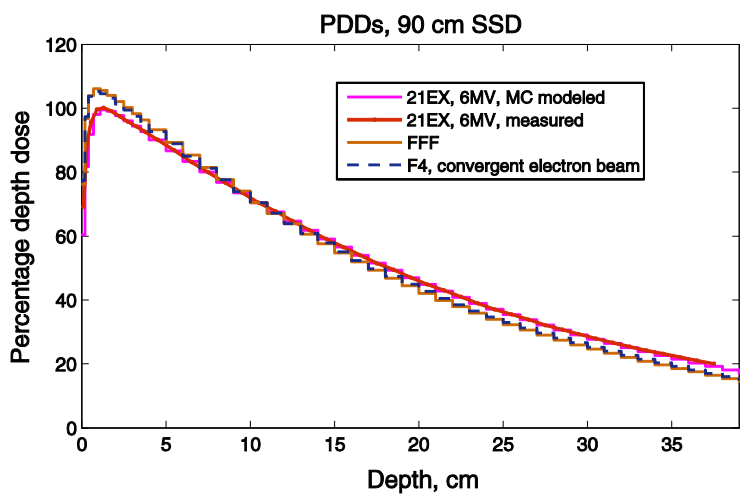

170

Figure 4. Percentage depth dose from for 40x40 cm$^2$ field calculated for conventional FF, FFF and F4 beams. Measured PDD from 21EX Varian linac is also shown. All PDDs were normalized at the depth of 10 cm.

175 Figure 4 shows PDD for 40 x 40 cm$^2$ field from conventional FF beam as compared to measured PDD as well as PDDs of FFF and F4 beams. This figure again shows excellent agreement of modeled PDD with the measurement that is known to be difficult to model for this large field. These results confirm validity of our 21EX model. Figure 4 also demonstrates that PDDs for FFF and F4 fields are very similar (F4 is slightly harder) and both less penetrating than standard FF beam.

### Optimal angular distribution of electrons, incident on the target, to produce F4 beam

Seven, out of twenty one calculated diagonal profiles that were used in optimization of the angular distribution of incident electrons, are shown in Figure 5. This figure shows that as convergence angle $\theta$ increases, the dose profile flattens but the flattening saturates once convergence angle reaches $15^0$. Expectedly, the dose rate also drops (Figure 6) with increase of convergence angle because the peak of the bremsstruhlang lobe diverges from the central axis of the beam and increasing fraction of the lobe gets absorbed by the primary collimator.

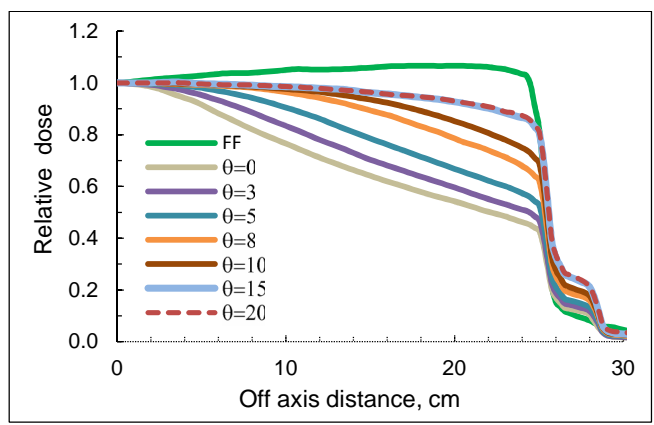

Figure 5. The dose profiles calculated by modeling seven narrow ($1^0$) conical beams (geometry shown in Figure 2) with convergence angles from $0^0$ to $20^0$. Profile from the standard FF beam is shown for comparison.

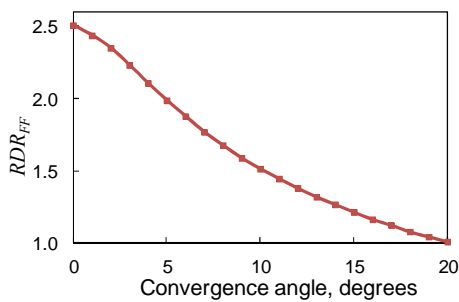

Figure 6. Dose rates, relative to FF beam, calculated for narrow ($1^0$) conical convergent beams

With flatness improving with increase of convergence angle it is expected that higher convergence angles provide flatter dose. However, with the dose rate being a boundary condition of optimization and dropping with increase of convergence angle, the algorithm is searching for a combination of angles that provides flattest summary profile while maintaining required dose rate. Therefore some forward directed component of the electron beam can be expected to maintain the dose rate. Optimization results showed that where the dose rate over about 1.5 times of $RDR_{FF}$, was requested, the angular distribution contained a forward (zero angle) component as well as oblique convergence angles. However, if lower dose rate was used as the boundary condition, a narrow distribution of oblique angles provided optimal solution. Solution space was found to be rather broad with multiple distributions providing similar profiles. In fact oblique angles only, with no forward component, provided near-optimal distributions as well. Optimized angular electron distributions for a set of five $RDR_{min}$ values are shown in Figure 7.

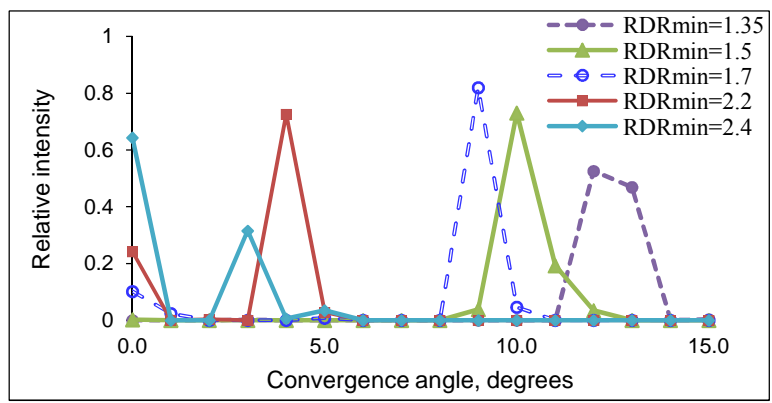

Figure 7. Distributions of electron convergence angles that provided flattest dose profiles for a set of five $RDR_{min}$ values shown in the legend.

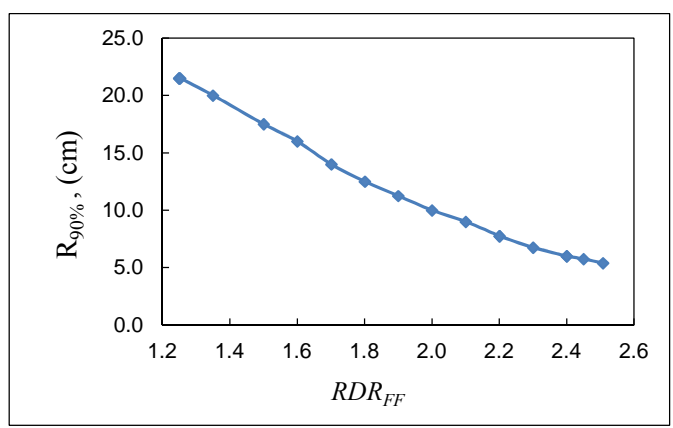

Figure 8. The radii of 90% dose levels in the optimized diagonal dose profiles as a function of achievable relative dose rate $RDR_{FF}$.

The radii of 90% dose level in the optimized dose profiles $D_{opt}(r)$ are shown in Figure 8. As expected, these radii reduce with increase of achievable dose rate. However the plot shows that the field size as large as 44 cm in diameter is potentially achievable with the dose rate 25% higher than that of FF linac. The field of 24 cm diameter, that covers majority of the modern conformal filed sizes, could potentially be achieved with the dose rate that is 80% higher than that of FF fields. Larger fields, apart from increased dose rates, would still have the benefits associated with the absence of flattening filter in these beams.

**Characteristics of F4 beams as compared to FF and FFF beams.**

Figure 9 shows diagonal profiles at the depth of maximum dose ($d_{max}$) as well as the depth of 10 cm modeled for FF, FFF and F4 beams. These profiles are normalized at the beam central axis and demonstrate that convergent electrons greatly improved flatness of the FFF beam. For example, for profiles calculated at 20 cm off-axis the dose increased from 52% for FFF beam to 92% for F4. At 15 cm off-axis the dose increased from 62% to 96%. In fact at $d_{max}$ for the area within the radius of 18 cm F4 beam is actually <u>flatter</u> than FF beam.

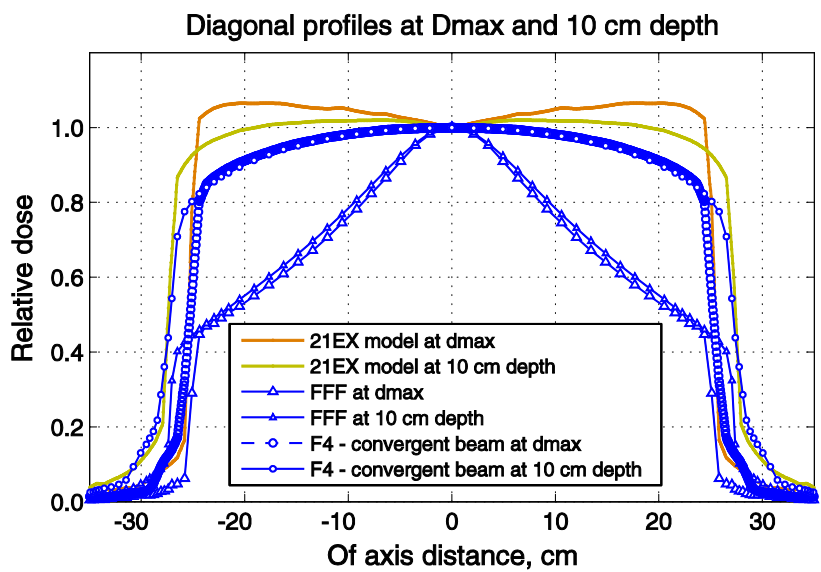

225    Figure 9. Diagonal dose profiles in water calculated using our MC model of 21EX linac for open 40x40 cm² 6MV FF beam, open 40x40 cm² 6MV FFF beam, open 40x40 cm² 6MV FFF beam *and* a convergent electron beam incident on the same target (F4 beam). The profiles were calculated for the water phantom positioned at 90 cm SSD and taken at the depth of $d_{max}$ (1.5 cm) as well as 10 cm.

Figure 9 also shows diagonal profiles for these beams at the depth of 10 cm. This figure demonstrates
230    that profiles of F4 and FFF beams do not change considerably with depth, unlike profiles of conventional FF beams. This has been previously reported for FFF beams (Kragl, af Wetterstedt *et al.*, 2009), and is expected, as very little off-axis photon spectral variation exists in these beams (Dalaryd, Kragl *et al.*, 2010).

Also, compared to FF beam, the out-of-field dose (Figure 10) was reduced by about a factor of two for F4
235    beam as it was for FFF beam. This dose reduction has been previously reported for FFF beams (Kry and et al., 2010, Almberg, Frengen *et al.*, 2012, Cashmore, Golubev *et al.*, 2012) and Figure 10 confirms that this also maintains for F4 beams.

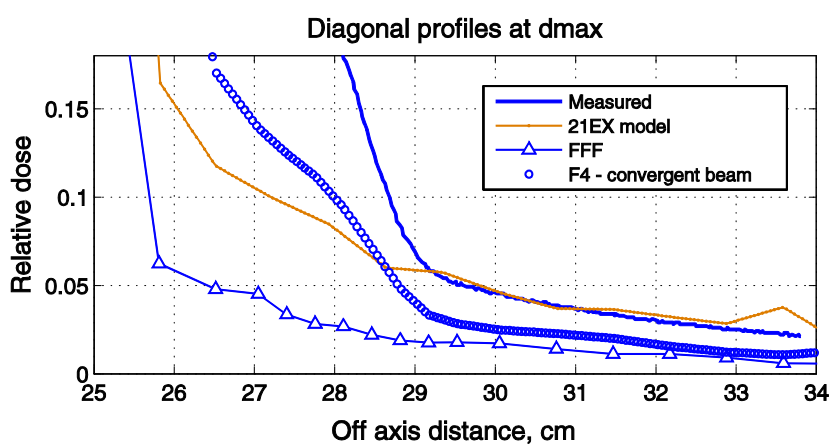

Figure 10. Off-axis fraction of 40x40 cm² diagonal profiles demonstrate reduction of out-of-field dose by
240    about a factor of two for FFF and F4 beams as compared to FF beam from 21EX model.

Figures 11-13 show the dose profiles calculated at the depths of $d_{max}$ as well as 10 cm for 4 x 4, 10 x 10 and 20x 20 cm² fields of FF and F4 beams. For the small beams of 4 x 4 cm² there is no considerable difference between FF and F4 profiles. For 10 x 10 cm² fields F4 profiles at both depths are similar to FF profiles at 10 cm depth and would probably perform as well in clinical, including IMRT, situations. For 245    20 x 20 cm² fields F4 profiles are ~2-3% lower than FF profile at 10 cm depth but still rather flat and most likely to be well acceptable for IMRT and VMAT treatments.

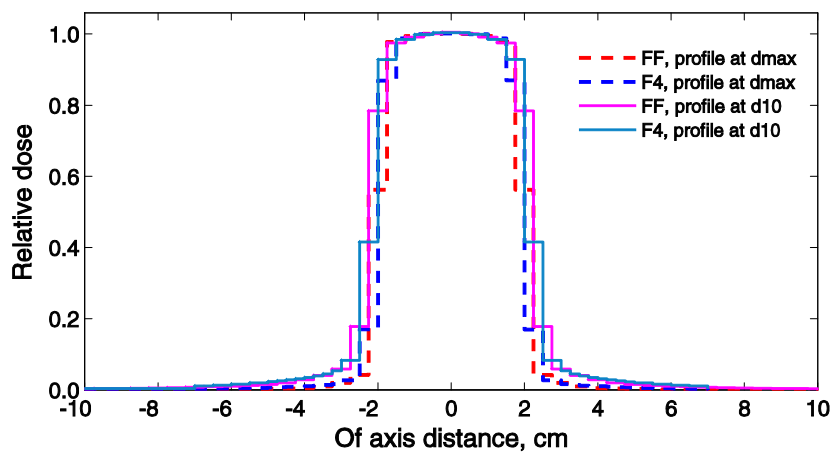

Figure 11. 6MV, 4x4 cm² dose profiles at dmax as well as the depth of 10 cm (d10) calculated for conventional FF and F4 beams. All profiles are normalized at CAX.

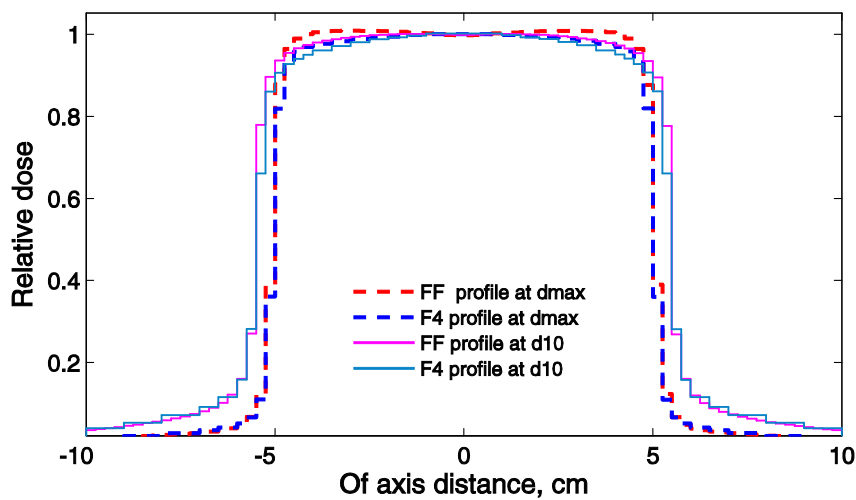

250

Figure 12. 6MV, 10x10 cm² dose profiles at dmax as well as the depth of 10 cm (d10) calculated for conventional FF and F4 beams. All profiles are normalized at CAX.

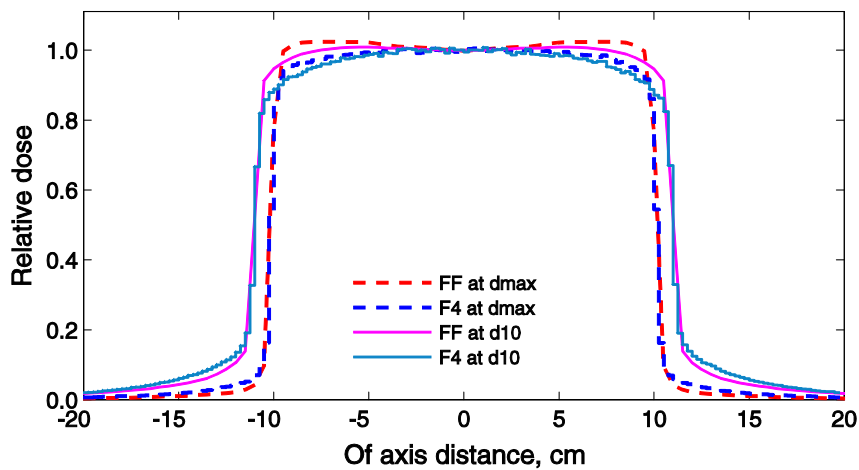

Figure 13. 6MV, 20x20 cm$^2$ dose profiles at dmax as well as the depth of 10 cm (d10) calculated for conventional FF and F4 beams. All profiles are normalized at CAX.

Finally, Figure 14 shows depth dose curves calculated for FF and F4 beams with different field sizes ranging from 4 x 4 to 40 x 40 cm$^2$. All F4 PDDs are less penetrating compared to PDDs from conventional 6MV FF beam. They also produce steeper build-up and higher surface dose, as expected from the softer beams.

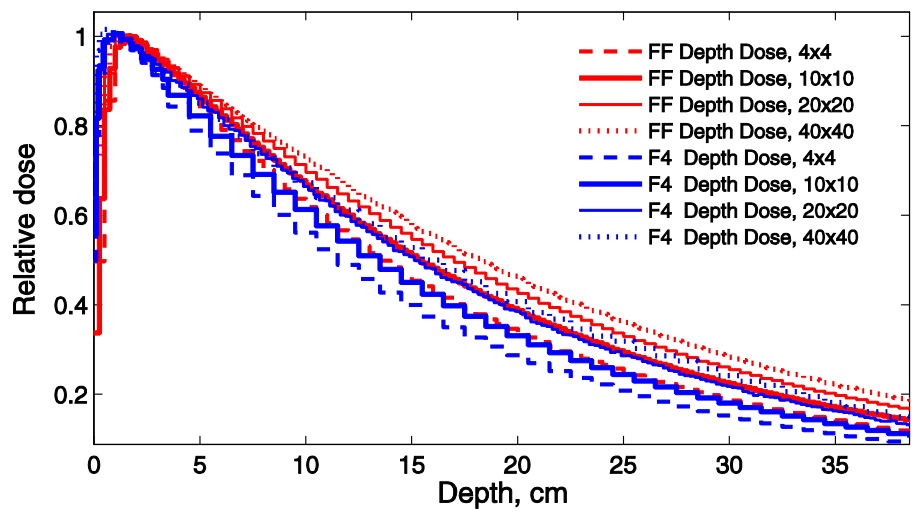

Figure 14. PDDs for conventional FF and F4 6 MV beams and field sizes from 4x4 to 40x40 cm$^2$. All profiles are normalized at the depth of 1.5 cm.

## Discussion

This paper, using Monte Carlo calculations combined with simulated annealing optimization, evaluated angular distributions of incident electrons that are capable of providing flat dose profiles, from FFF accelerator. The beam collimation system of Varian 21EX linac was used in this study for the sake of simplicity; however different linac model and beam energies would produce qualitatively similar results. This study shows that it is possible to achieve the dose profile at dmax that is about as flat as profile of conventional FF beam (though the dose off - axis is reduced rather than increased as it is for FF beam), and the flatness is not much reduced at the depth of 10cm.

Although all data in this study have been produced using convergent electrons incident on the target, divergent beams have been also modeled and the results were very similar to those presented in this paper. Divergent beams could be modeled by simply moving the beam convergence point a few millimeters above the target. As electrons cross over convergence point, divergent beam will enter the target. Clearly, characteristics of such beam will be very similar to those presented here and therefore were not shown in this paper.

Electron beams, modeled in this paper, provided flatter fields than those achievable from de-focusing of the electrons incident on the target (Tsiamas, Seco *et al.*, 2011). The dose profiles shown by Tsiamas et al (only profiles for 20x20cm$^2$ filed size are shown, see Fig. 4, (Tsiamas, Seco *et al.*, 2011)) indicate that for 20x20 cm$^2$ field the flattest profile that was produced by 14$^0$ electron beam angular spread and resulted in ~92% dose at 8.5cm off axis. Our 20x20cm$^2$ profile (Figure 9) provided 95% of the CAX dose demonstrating more efficient flattening capabilities of the beams modeled in this study. Our study also shows the feasibility of achieving +/-5% flat dose profiles within 22cm off axis from 40x40 cm$^2$ FFF fileds.

Figure 10 shows that 40x40 cm$^2$ F4 beam considerably reduced scattered radiation from the treatment head compared to FF: phenomenon which has been previously reported for FFF beams. In general, the properties of F4 beams such as softer x-ray spectrum with less penetrating PDDs, reduced leakage and scattered radiation are very similar to those of FFF beams. This is expected, as geometries of these beams are very similar, indeed. In addition, similarly to that of FFF beams, F4 beams have near no spectral variability of the beam off-axis which reduces variation in field-size dependent dosimetric parameters. This should simplify modeling of such beams within treatment planning systems resulting in more accurate dose calculations.

As convergence angle of incident electron beam increases, increasingly larger fraction of the forward-peaked bremsstrahlung lobe hits the primary collimator and gets absorbed resulting in reduced dose rate. The size of flat area of the beam therefore becomes a function the maximum achievable dose rate (Figure 8), and it is foreseeable to utilize this effect and use different angular distributions of the incident electrons at different field size settings to maintain flat dose while producing the highest achievable dose rates.

Flat dose distributions have been previously achieved with racetrack accelerator that had no flattening filter through the use of scanning magnets as was proposed by Brahme et al (Brahme, Kraepelien *et al.*, 1980). The authors of this and subsequent papers on racetrack MM50 (Karlsson, Nystrom *et al.*, 1993, Satherberg and Karlsson, 1998) discussed scan patterns that produce uniform dose from 50 MeV incident electrons. In fact their pattern qualitatively agrees the optimal angular distributions derived in this paper. Similar to our forward and oblique components of the incident electron distribution (Figure 7), their scanning pattern also had central component as well as peripheral circular pattern (Figure 2 in Karlsson et al (Karlsson, Nystrom *et al.*, 1993)). However flat dose distributions were only achievable from MM50 when using high energy electrons. At the energies of 10 MV and for small fields at 20 MV a cone-shaped graphite absorber was used to provide extra beam flattening as well as absorption of primary electrons.

This paper therefore presents the first study investigating possible production and dosimetry of flat FFF beams from low (6MeV) energy electrons. Scanning electron beam technology could potentially be used to achieve convergent electron beams modeled in this paper. In order to produce convergent beams, investigated in this paper, the electrons incident on the target would have to be swept dynamically providing azimuthally uniform electron fluence. Detailed discussion on production of such beams is outside of the scope of this paper.

Dosimetric properties of F4 compared to FF beams would be most beneficial for treatments that require large fields such as head and neck IMRT and VMAT. F4 beams could also be used in conformal treatments. They provide higher efficiency with similar target coverage as FF beams and would not require as high number of monitor units for these treatments as FFF beams do. Lower, compared to FF beam, leakage, and scatter and out of field dose would also be expected.

## Conclusions

This preliminary research has shown potential for achieving flat FFF beams. Dosimetry of such beams has been evaluated. These beams are shown to combine dosimetric and clinical advantages of FF and FFF beams potentially offering more efficient treatments with less scatter, less leakage radiation and less monitor units required to deliver the dose.

## References:

Abacioglu M, Unal S, Ozen Z, *et al.* 2012 Head and Neck IMRT With or Without Flattening Filter: A Dosimetric Comparison *Int J Radiat Oncol Biol Phys* **84** S872-S73

Almberg S S, Frengen J and Lindmo T 2012 Monte Carlo study of in-field and out-of-field dose distributions from a linear accelerator operating with and without a flattening-filter *Med. Phys.* **39** 5194-203

Arcangeli S, Castiglioni S, Mancosu P, *et al.* 2012 Is Stereotactic Body Radiation Therapy (SBRT) an Attractive Option for Unresectable Liver Metastases? Early Results From a Phase 2 Trial *Int J Radiat Oncol Biol Phys* **84** S3

Atwood T F, Mok E, Lo A, *et al.* 2011 Dosimetric Comparison of RapidArc versus CyberKnife for Stereotactic Body Radiation Therapy for Pancreatic Cancer *Int J Radiat Oncol Biol Phys* **81** S345-S46

Basran P S, Zavgorodni S, Berrang T, *et al.* 2010 The impact of dose calculation algorithms on partial and whole breast radiation treatment plans *Radiat Oncol* 16 5:120

Brahme A, Kraepelien T and Svensson H 1980 Electron and photon beams from a 50 MeV racetrack microtron *Acta Radiol Oncol* 19 305-19

Bush K, Gagné I M, Zavgorodni S, *et al.* 2011 Dosimetric validation of Acuros® XB with Monte Carlo methods for photon dose calculations *Med. Phys.* 38 2208-21

Bush K, Townson R and Zavgorodni S 2008 Monte Carlo simulation of RapidArc radiotherapy delivery *Phys. Med. Biol.* N359-70

Bush K, Zavgorodni S and Beckham W 2009 Inference of the optimal pretarget electron beam parameters in a Monte Carlo virtual linac model through simulated annealing *Med. Phys.* 36 2309-19

Cashmore J 2008 The characterization of unflattened photon beams from a 6 MV linear accelerator *Phys. Med. Biol.* 1933

Cashmore J, Golubev S, Dumont J L, *et al.* 2012 Validation of a virtual source model for Monte Carlo dose calculations of a flattening filter free linac *Med. Phys.* 39 3262-69

Chibani O, Moftah B and Ma C M C 2011 On Monte Carlo modeling of megavoltage photon beams: A revisited study on the sensitivity of beam parameters *Medical Physics* 38 188-201

Chofor N, Harder D, Willborn K, *et al.* 2011 A direction-selective flattening filter for clinical photon beams. Monte Carlo evaluation of a new concept *Phys Med Biol* 56 4355-76

Dalaryd M, Kragl G, Ceberg C, *et al.* 2010 A Monte Carlo study of a flattening filter-free linear accelerator verified with measurements *Phys Med Biol* 55 7333

Duan J, Spencer S A, Shen S, *et al.* 2011 Clinical Performance of Automated IMRT with High Intensity Flattening-Filter-Free Beams in Comparison to VMAT for Head and Neck Cancers *Int J Radiat Oncol Biol Phys* 81 S548-S49

Dzierma Y, Licht N, Nuesken F, *et al.* 2012 Beam properties and stability of a flattening-filter free 7 MV beam---An overview *Med. Phys.* 39 2595-602

Gagné I M and Zavgorodni S F 2007 Evaluation of the analytical anisotropic algorithm (AAA) in an extreme water-lung interface phantom using Monte Carlo dose calculations *J. Appl. Clin. Med. Phys.* 8 33-46

Georg D, Knoos T and McClean B 2011 Current status and future perspective of flattening filter free photon beams *Med. Phys.* 38 1280-93

Hrbacek J, Lang S and Klöck S 2011 Commissioning of Photon Beams of a Flattening Filter-Free Linear Accelerator and the Accuracy of Beam Modeling Using an Anisotropic Analytical Algorithm *Int J Radiat Oncol Biol Phys* 80 1228-37

Karlsson M, Nystrom H and Svensson H 1993 Photon beam characteristics on the MM50 racetrack microtron and a new approach for beam quality determination *Medical Physics* 20 143-49

Kawrakow I, Rogers D W and Walters B R 2004 Large efficiency improvements in BEAMnrc using directional bremsstrahlung splitting *Med. Phys.* 31 2883-98

Kim G, Rice R, Lawson J, *et al.* 2012 Stereotactic Radiosurgery With FFF Mode Photon Beams *Int J Radiat Oncol Biol Phys* 84 S823

Kragl G, af Wetterstedt S, Knäusl B, *et al.* 2009 Dosimetric characteristics of 6 and 10MV unflattened photon beams *Radiother.Oncol.* 93 141-46

Kragl G, Albrich D and Georg D 2011 Radiation therapy with unflattened photon beams: Dosimetric accuracy of advanced dose calculation algorithms *Radiother.Oncol.* 100 417-23

Kry S F and et al. 2010 Out-of-field photon dose following removal of the flattening filter from a medical accelerator *Phys Med Biol* 55 2155

Navarria P, Pentimalli S, Ascolese A, *et al.* 2012 Lung Metastases in Oligometastatic Patients: A New SBRT Approach Using VMAT With Flattening Filter-free (FFF) Beams *Int J Radiat Oncol Biol Phys* 84 S579

Nicolini G, Ghosh-Laskar S, Shrivastava S K, *et al.* 2012 Volumetric Modulation Arc Radiotherapy With Flattening Filter-Free Beams Compared With Static Gantry IMRT and 3D Conformal Radiotherapy for Advanced Esophageal Cancer: A Feasibility Study *Int J Radiat Oncol Biol Phys* 84 553-60

Park J M, Kim J-i, Choi C H, *et al.* 2012 Photon energy-modulated radiotherapy: Monte Carlo simulation and treatment planning study *Medical Physics* 39 1265-77

Ponisch F, Titt U, Vassiliev O N, *et al.* 2006 Properties of unflattened photon beams shaped by a multileaf collimator *Med. Phys.* 33 1738-46

Rogers D W O, Walters B R and Kawrakow I. 2009. *BEAMnrc Users Manual*. NRCC Report PIRS-0509(A) rev K, National Research Council of Canada, Ottawa.

Satherberg A and Karlsson M 1998 Calculation of photon energy and dose distributions in a 50 MV scanned photon beam for different target configurations and scan patterns *Medical Physics* 25 236-40

St-Hilaire J, Sevigny C, Beaulieu F, *et al.* 2009 Optimization of photon beam energy in aperture-based inverse planning *J Appl Clin Med Phys* 10 3012

Stevens S W, Rosser K E and Bedford J L 2011 A 4 MV flattening filter-free beam: commissioning and application to conformal therapy and volumetric modulated arc therapy *Phys Med Biol* 56 3809

Thomas E, Popple R A, Prendergast B M, *et al.* 2012 Improved Treatment Efficiency With Flattening Filter-free Mode and Implications for Common Clinical Scenarios *Int J Radiat Oncol Biol Phys* 84 S860

Titt U, Vassiliev O N, Ponisch F, *et al.* 2006a A flattening filter free photon treatment concept evaluation with Monte Carlo *Med. Phys.* 33 1595-602

Titt U, Vassiliev O N, Ponisch F, *et al.* 2006b Monte Carlo study of backscatter in a flattening filter free clinical accelerator *Med. Phys.* 33 3270-73

Tsiamas P, Seco J, Han Z, *et al.* 2011 A modification of flattening filter free linac for IMRT *Med. Phys.* 38 2342-52

Vassiliev O N, Kry S F, Kuban D A, *et al.* 2007 Treatment-Planning Study of Prostate Cancer Intensity-Modulated Radiotherapy With a Varian Clinac Operated Without a Flattening Filter *Int. J. Radiat. Oncol. Biol. Phys.* 68 1567-71

Vassiliev O N, Titt U, Kry S, F, *et al.* 2006 Monte Carlo study of photon fields from a flattening filter-free clinical accelerator *Med. Phys.* **33** 820-27

Verbakel W, Ong C, Senan S, *et al.* 2012 Flattening Filter-free Beams for SBRT: Advantages and Risks *Int J Radiat Oncol Biol Phys* **84** S826-S27

Wu Q, Yoo S, Das S, *et al.* 2012 Evaluation of IMRT and VMAT for Head-and-Neck Cancer With Flattening Filter-free (FFF) Beams *Int J Radiat Oncol Biol Phys* **84** S863

Xie J, Cao J, Chen W, *et al.* 2012 Breast Tangential Treatment Plan Using Flattening Filter-free Beams: A Planning Study *Int J Radiat Oncol Biol Phys* **84** S867

Zavgorodni S, Bush K, Locke C, *et al.* 2007 Vancouver Island Monte Carlo (VIMC) system for radiotherapy treatment planning dosimetry and research *Radiother. Oncol.* **84** Supplement 1, S49

Zavgorodni S, Locke C, Bush K, *et al.* 2005 Correcting the BEAMnrc phase space to match Monte Carlo and measured dose distributions *Australas. Phys. Eng. Sci. Med.* **28** 285

Zhang G G, Ku L, Dilling T J, *et al.* 2011 Volumetric modulated arc planning for lung stereotactic body radiotherapy using conventional and unflattened photon beams: a dosimetric comparison with 3D technique *Radiat Oncol* **6** 152

Zhu X R, Kang Y and Gillin M T 2006 Measurements of in-air output ratios for a linear accelerator with and without the flattening filter *Med. Phys.* **33** 3723-33

Zwahlen D R, Lang S, Hrbacek J, *et al.* 2012 The Use of Photon Beams of a Flattening Filter-free Linear Accelerator for Hypofractionated Volumetric Modulated Arc Therapy in Localized Prostate Cancer *Int J Radiat Oncol Biol Phys* **83** 1655-60